\documentclass{eas}
\usepackage{graphicx}

\usepackage{amsmath}
\usepackage{natbib}
\bibpunct{(}{)}{;}{a}{}{;}

\TitreGlobal{The Title of this Volume}

\begin{document}
\title{NLTE water lines in Betelgeuse-like atmospheres} 
\author{Julien Lambert}\address{LUPM â UMR5299, CNRS \& Universit\'e Montpellier II, Place Eug\`ene Bataillon, 34095 Montpellier Cedex 05, France \email{Julien.Lambert@astro.lu.se}}
\secondaddress{Lund Observatory, Box 43, SE-22100 Lund, Sweden}
\author{Eric Josselin}\sameaddress{1}
\author{Nils Ryde}\sameaddress{2}
\author{Alexandre Faure}\address{IPAG - UMR5274, CNRS \& UJF-Grenoble, F-38041 Grenoble, France}
\begin{abstract}
The interpretation of water lines in red supergiant stellar atmospheres has been much debated over the past decade. The introduction of the so-called MOLspheres to account for near-infrared "extra" absorption has been controversial. We propose that non-LTE effects should be taken into account before considering any extra-photospheric contribution. 

After a brief introduction on the radiative transfer treatment 
and the inadequacy of classical treatments in the case of large-scale systems such as molecules, we present a new code, based on preconditioned Krylov subspace methods. Preliminary results suggest that NLTE effects lead to deeper water bands, as well as extra cooling. 
\end{abstract}
\maketitle

\section{Water in RSG atmospheres}
The interpretation of water lines in red supergiant stellar atmospheres has been debated 
for more than a decade. For example, \cite{tsuji2000} shows that near-infrared water bands 
observed with the Infrared Satellite Observatory (ISO) low resolution spectrograph 
toward $\mu$ Cep appear significantly deeper than expected from a pure photospheric, 
1D, local thermodynamic equilibrium (LTE) and hydrostatic synthetic spectrum. In order to 
quantify this extra absorption, he  considered the possibility to add an {\it ad hoc} quasi-static 
molecular shell baptized "MOLsphere". But what was initially introduced as toy model gained 
support through interferometric observations which reveal a larger spatial extension in water 
bands compared to the stellar radius determined in the continuum \citep[see][]{perrin2005, ohnaka2004}. 

However the MOLsphere model is still subject to some weaknesses. First of all, 
no model is up to now able to explain the formation of such extra-photospheric layer 
(even gas levitation by shock wave cannot account  for the parameters such as density and water 
abundance proposed in MOLsphere models). 
One should also keep in mind that, due to the rather poor coverage of the Fourier plane 
by current infrared interferometers, the interpretation of such observations strongly rely on a 
presupposed model. More importantly, based on high resolution spectra 
obtained with TEXES, \cite{ryde2006} showed that water lines around 12 $\mu$m, predicted in 
emission by MOLsphere models, are in fact observed in absorption and are probably of photospheric origin.

In order to reconcile the interpretation of observations in various spectral ranges, 
we want to test how important non LTE (NLTE) effects are and whether they may mimic a MOLsphere 
through a strengthening of absorption lines. 
These NLTE effects may also affect the thermodynamical structure of the atmospheres 
through e.g. the contribution of molecules to the cooling function and the radiative pressure. 
If it leads to a more extended atmosphere and a shift of the water line contribution function, 
this could potentially explain the interferometric observations. 
\section{General context of radiative transfer in molecular lines}
\subsection{Basics of radiative transfer}
In order to determine the specific intensity $I_\nu$ in a line, one has to solve the radiative transfer equation 
(RTE). For a 1D plane parallel geometry, this equation is a simple ordinary differential equation of the first order, 
\begin{equation}
 \mu\frac{dI_{\nu}}{d\tau_{\nu}}=I_{\nu}-S_{\nu} .
\end{equation}
Solving the RTE under NLTE conditions is however not straightforward, because of an intricate 
non-linear and non-local coupling between level populations and radiation. 
Indeed the determination of the  source function for a given line due to a transition 
$i \rightarrow j$ at an optical depth $\tau$ requires the knowledge of the populations on the 
levels $i$ and $j$ at this optical depth.
\begin{equation}
 S_{\nu_{ij}}(\tau)=\frac{2h{\nu_{ij}}^3}{c^2}\frac{1}
   {\frac{x_j(\tau)}{x_l(\tau)}}-1 ,
\end{equation}
where $x_k = n_k/g_k$, with $n_k$ being the relative population of level $k$ and $g_k$ is its statistical weight. The other symbols have their usual meanings. 
These populations can be derived by solving the statistical equilibrium system of equations, which is a set of rate equations for each energy level.
\begin{equation}
 \begin{split}
   n_i&\left[\sum^N_{j<i}{A_{ij}+\sum^N_{j\ne i}{\left(B_{ij}{{\overline{\mathcal{J}}}}_{ij}
   + n_{col}C_{ij}\left(T\right)\right)}}\right]\\
   -&\left(\sum^N_{j>i}{n_jA_{ji}+\sum^N_{j\ne i}{n_j\left(B_{ji}{{\overline{\mathcal{J}}}}_{ji}
   + n_{col}C_{ji}\left(T\right)\right)}}\right)=0 , 
    \end{split}
\end{equation}
where ${\overline{\mathcal{J}}}_{ji}$ denotes the mean radiation field for the $i \rightarrow j$ transition, integrated over 
the line profile. 
This system has in general no analytical solution. However, if one considers that the probability of radiative 
(de)excitation is very small compared to the collisional (de)excitation, either because the radiative field 
(through  ${\overline{\mathcal{J}}}_{ji}$) is weak or because the density of colliders 
$n_{col}$ is large, the system can be simplified and leads to an analytical solution for the level populations, being the Boltzmann distribution. 
The populations then only depend on the local kinetic temperature at $\tau$. This is the LTE hypothesis. 
The source function is simplified to the Planck function which only depends on the \emph{local} temperature as well. 
On the contrary, if the radiative field is strong and/or the collider density is low, one has to consider the radiative 
terms associated to ${{\overline{\mathcal{J}}}}$ and invert this system of equations. The main difficulty is that the 
mean intensity field ${{\overline{\mathcal{J}}}}$ depends on the solution of the RTE:
 \begin{equation}%
 \label{barj}
    {\overline{\mathcal{J}}}_{ij}({\tau }) = \frac{1}{2} \int_{0}^{\infty}{ \phi_\nu (\tau)%
      \int_{-1}^{+1} {I_\nu(\tau,\mu) d\mu} d\nu}%
  \end{equation}%
This explicit dependence on the solution means that to solve the statistical equilibrium in order to 
determine the source function, one need to know the intensity at different locations and for different 
frequencies simultaneously. 
The radiative transfer is thus a non-local problem where the physical state at a given $\tau$ is coupled with every other layer. Furthermore, for a multi-level system, the problem is also non-linear. Because of this coupling, the source function and the mean intensity field are usually determined iteratively. This is the classical way to solve NLTE radiative transfer.

\runningtitle{short title}

\subsection{The water molecule}
\begin{figure}[ht!]
\centering
\includegraphics[width=\textwidth]{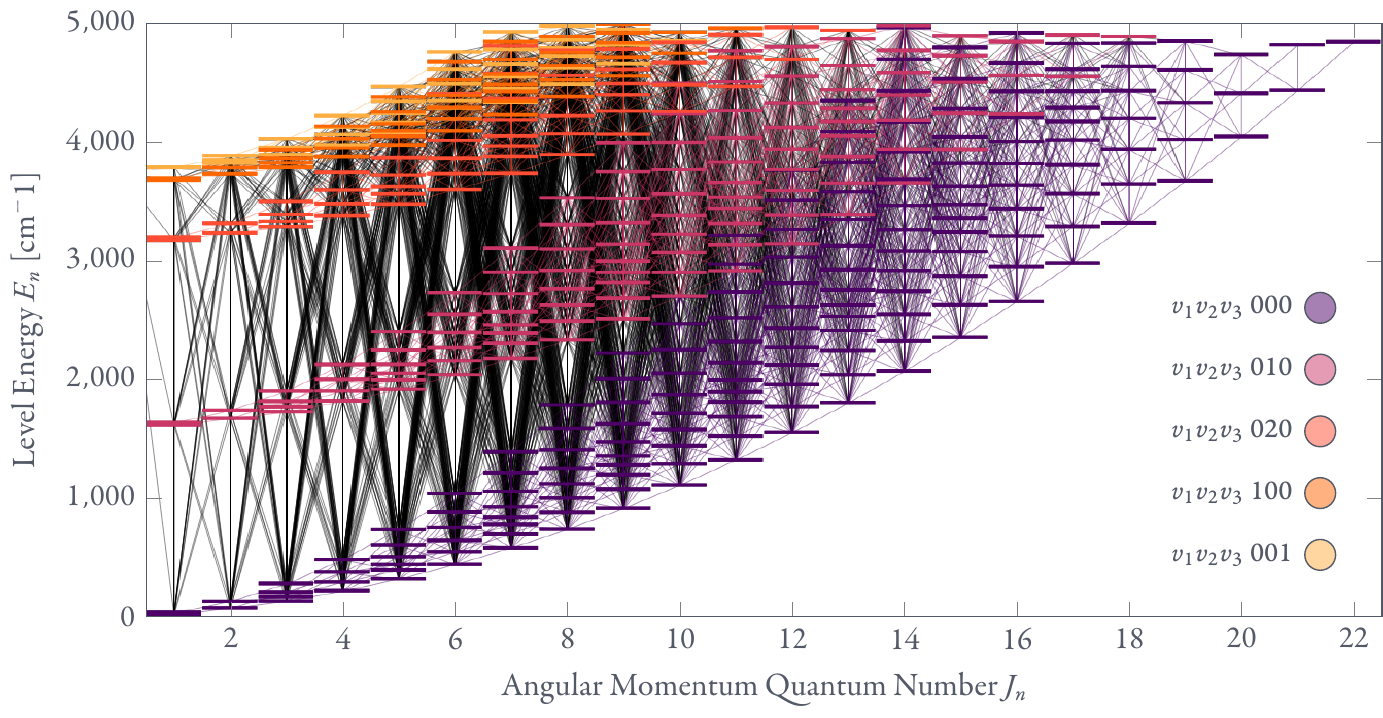}
\caption{Grotrian diagram of the water molecule with the radiative transitions taken into account in this work. $J_n$ is the total angular momentum.}
\label{grotrian}
\end{figure}

Solving the statistical equilibrium system requires molecular data: the energy and statistical weight of each level, the Einstein coefficients $A_{ij}$  (including the selection rules), and finally, the collisional rates $C_{ij}$. For the water 
molecule, we have used the line list and radiative coefficients from \cite{barber2006}. The radiative transitions we have 
taken into account are displayed in Fig. \ref{grotrian}. 
The collisional rates were computed by \cite{faure2008} thanks to an original method of extrapolation for high $J$ (relevant to the temperatures met in red supergiant -  RSG -  atmospheres).

A first rough estimate of the deviation from LTE can be obtained by evaluating the competition between radiative and collisional processes. A good indicator of this deviation is thus the ratio between radiative and collisional de-excitation rates, that is, the ratio between the critical density and the density of colliders for each energy level. 
\begin{equation}
    \eta_i(\tau)=\frac{n^{cri}_i(\tau)}{n^{col}(\tau)}=
    \frac{1}{n^{col}(\tau)}\frac{\sum^{i-1}_{j=1}{A_{ij}}}
      {\sum^{i-1}_{j=1}{C_{ij}(T(\tau))}}
\end{equation}
If this ratio $\eta_i$ is much smaller than 1 for every level $i$, collisions 
dominate everywhere over radiative transitions and the solution of the statistical equilibrium equations tends toward the Boltzmann distribution (LTE). On the contrary, if $\eta_i$ is greater than one, NLTE conditions are met. 
\begin{figure}[ht!]
\centering
\includegraphics[width=\textwidth]{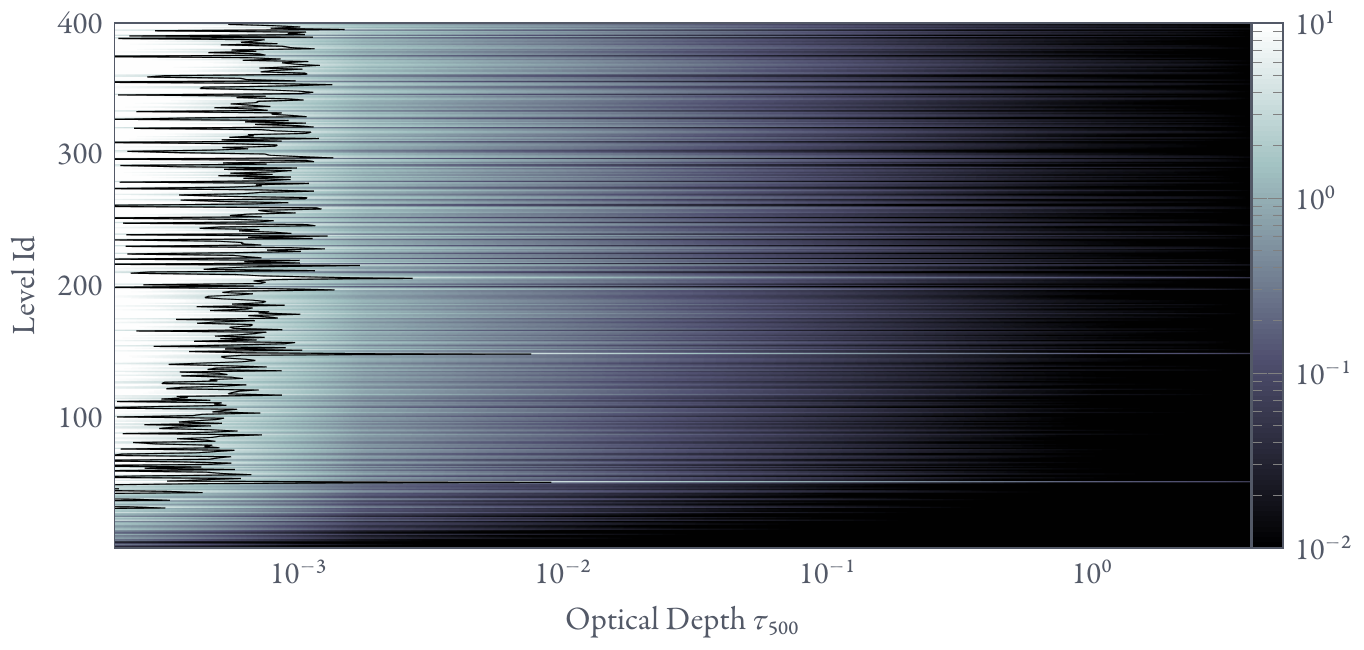}
\caption{Ratio between the critical density and the density of colliders (e$^-$ and H$_2$) in a Betelgeuse-like atmosphere. The solid line indicates the transition  $\eta_i = 1$.}
\label{critical}
\end{figure}

This ratio is computed and displayed in Fig. \ref{critical} in the case of water in a typical RSG atmosphere. $\eta_i$  is much larger than 1 in the upper atmosphere, i.e. where the water line contribution function is expected to peak. This results mostly from the 
water abundance profile and the excitation temperature.   
An accurate NLTE treatment of the RTE is thus required to properly analyze water spectra. 

\subsection{The weakness of classical methods for large-scale systems}
\begin{figure}[ht!]
\centering
\includegraphics[width=\textwidth]{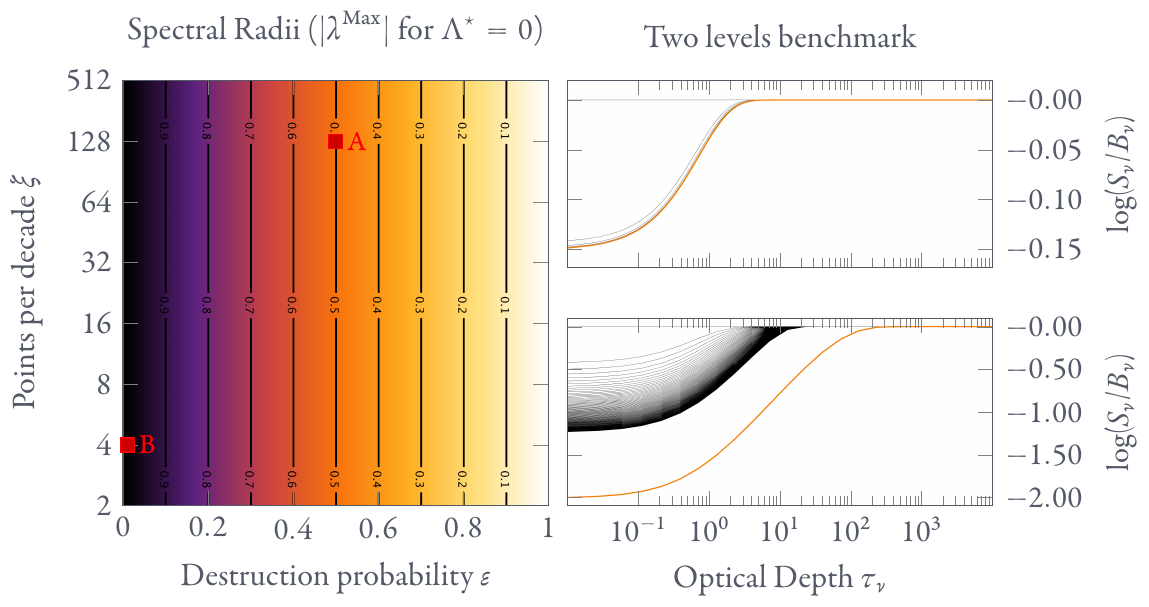}
\includegraphics[width=\textwidth]{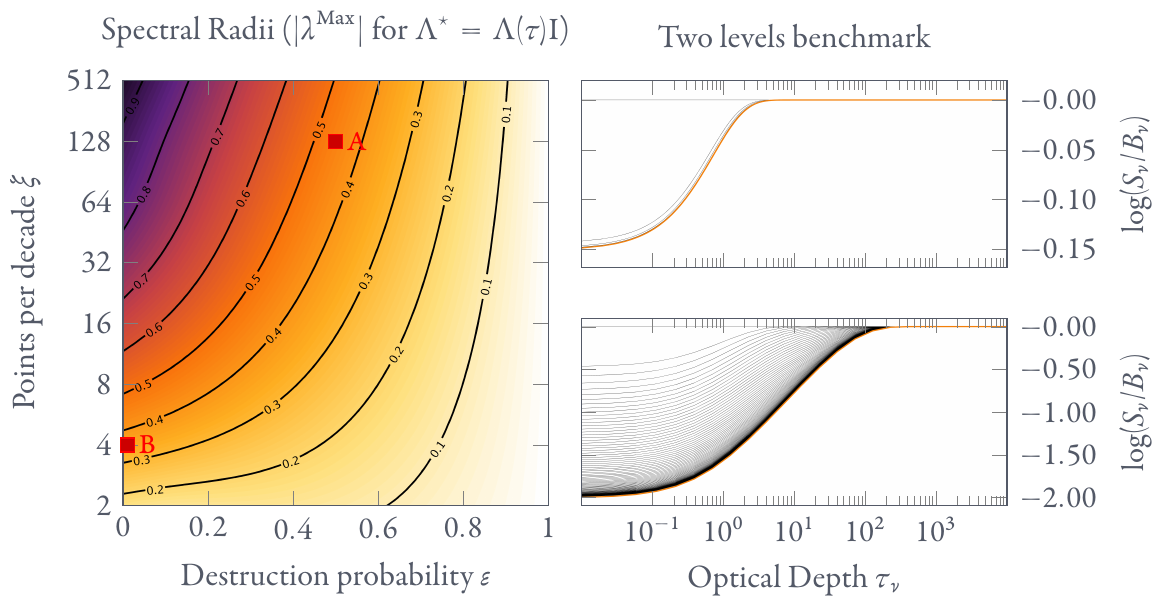}
\caption{Spectral radius (left) and convergence properties (right) for two static iterative methods (LI: upper part; ALI with 
$\Lambda^\star$ = diagonal of $\Lambda$: lower part). 
For each method, the convergence  is shown for two spectral radii, marked with letters A (upper panel) and B (lower panel) 
in the right-hand panels (the true solution is given there by the orange line). }
\label{radius}
\end{figure}

The classical methods generally used to solve the RTE rely on the operator splitting technique, which is derived from the Lambda Iteration (LI) \citep{trujillo-bueno1995}. The LI method consists in iteratively 
computing the Schwarzschild-Milne integral with the use of an integration operator 
$\Lambda$. $\Lambda$ integrates the source function over the domain (i.e. the stellar atmosphere) in order to determine the mean intensity field. However, LI is inefficient and inaccurate in  many cases, as it does not converge toward the true solution. This is due to the fact that the spectral radius of the amplification matrix of this scheme $\lambda^{max}$ 
depends on the NLTE parameter $\epsilon$ (i.e. the photon destruction probability; the lower $\epsilon$, the stronger the NLTE effects) and the total optical depth of the 
medium $T$, as $\lambda^{max}\simeq (1-\epsilon)(1-T^{-1})^{-1}$. It is close to one if $\epsilon$ is small (i.e. strongly NLTE) 
and/or if $T$ is large.
In order to improve its convergence toward the true solution, a better conditioning can be obtained through 
operator splitting methods, such as ALI (Jacobi), Gauss-Seidel or Successive OverRelaxation method SOR 
\citep{trujillo-bueno1995}, which reduce the spectral radius (see Fig. \ref{radius}). 
Unfortunately, for large discretized and strongly NLTE problems, the spectral radius remains large, and the convergence slow (e.g. typically $\sim$ 100 iterations in case B in Fig. \ref{radius}). This slow convergence becomes problematic if the model atom or molecule is large, as for each iteration the large statistical equilibrium system has to be solved. For water, this system is huge (about 1000 energy levels, corresponding to about 15,000 radiative rates and more than 330,000 collisional rates; see Fig. \ref{grotrian}) and the total time needed to converge becomes prohibitive. Furthermore, the accuracy of these methods may be poor \citep[see][especially Fig. 9]{chevallier2003}. This is intrinsically due to the static nature of the iterations.
\section{New method and preliminary results}
\subsection{Methodology}
Rather than determining iteratively the source function, the new method described in \cite{lambert2012} is to determine, in a unique step, the RTE and the populations in every layer of a discretised model atmosphere. This is done by inverting a generalised multi-zone statistical equilibrium system with a non-linear solver based on non-static methods such as sub-space Krylov methods 
\citep{Saad2003}.

The first step consists in rewriting the RTE with an explicit dependence on populations which become the new unknowns. Moreover, because of the non-locality of the RTE and the need to couple the sets of statistical equilibrium in every layer, we use a unique transition-independent scale (e.g. $\tau_\text{V}=\tau_\text{500}$).
\begin{equation}
    \mu\frac{dI_{\nu}}{d\tau_\text{V}}=\widetilde{\alpha}_\nu\left(x_j,x_i\right)I_{\nu} - \widetilde{j}_{\nu}\left(x_i\right)
\end{equation}
Then, the formal solution of this equation can be determined by the use of its Green function and its integration over the model atmosphere. By including this solution into Eq. \ref{barj}, an analytical form of the mean intensity is obtained, with an explicit dependence of the population:
\begin{equation}
    {\overline{J}}_{ij}({\tau }_{\text{V}})
    =\int^{{\tau }^\text{out}_\text{V}}_{{\tau }^\text{in}_\text{V}}
    {f_{ij}\left(x_i(t)\right) \cdot K_{ij}\left[
    x_i,x_j\right](t,\tau_\text{V}) dt}
\end{equation}
where $f_{ij}\left(x_i(t)\right)$ denotes the explicit dependence  on populations \citep[see][for more details]{lambert2012} and 
$K_{ij}$ is the kernel of the RTE. 
This expression is included into the statistical equilibrium equations of each layer (labelled by $k$, $1 \le k \le P$) and is 
rewritten in a matrix form.
\begin{equation}
    \textbf{M}^k\left[\bar{\textbf{J}}^k(\textbf{x}^1,\textbf{x}^2,...,\textbf{x}^P)\right]\cdot\textbf{x}^k=\textbf{0}
\end{equation}
Because the RTE is non-local, $\bar{\textbf{J}}^k$ requires an integral over ${\tau }_{\text{V}}$, and thus the knowledge of the populations everywhere $(\textbf{x}^1,\textbf{x}^2,...,\textbf{x}^P)$.
The well-adapted unknown vector is thus $\textbf{X}=(\textbf{x}^1,\textbf{x}^2,...,\textbf{x}^P)$ and the full closed system is built:
\begin{equation}%
    \label{eq:RTE6}%
    \boldsymbol{\Gamma}(\mathbf{X})\cdot\mathbf{X}=
     \left( \begin{array}{ccc}%
        {{\mathbf M}}^1\left({\mathbf X}\right) & 0 & 0\\%
        0 & \ddots  & 0\\%
        0 & 0 & {{\mathbf M}}^p\left({\mathbf X}\right) \end{array}%
    \right).%
    \left( \begin{array}{c}%
        \mathbf{x}^1\\%
        \vdots \\%
        \mathbf{x}^P%
      \end{array} \right)=%
    \left( \begin{array}{c}%
        \mathbf{0} \\%
        \vdots \\%
        \mathbf{0}%
      \end{array} \right)
  \end{equation}%
Solving the RTE and the statistical equilibrium equations then consists in finding the root of:
\begin{equation}%
    \label{eq:fonction}%
    \boldsymbol{\Gamma}(\mathbf{X})\cdot\mathbf{X}=\mathbf{F}(\mathbf{X})=\mathbf{0}%
  \end{equation}%
The inversion of this system gives the populations everywhere in the model which enables the computation of the source function. The emergent spectrum and any other physical quantity are easily post-processed.
An interesting advantage of this approach is the analytical form of this $N \times P$ integral equation (for $N$ energy levels and 
$P$ atmosphere layers), and thus the possibility to differentiate each equation with respect to populations on every level and 
every layer in order to build an analytical Jacobian matrix (which is sparse) very rapidly. 
\subsection{A new MPI code: MOrad}
This method has been implemented in a fully parallelised MPI code which takes as input a model atmosphere and molecular data. The chosen MPI strategy is a domain decomposition on spatial cells. The code computes the multi-D function $\mathbf{F}(\mathbf{X})$ and its Jacobian Matrix. It is interfaced with the efficient non-linear and \emph{scalable} library PETSc, which solves the equation with the "generalized minimal residual method" (GMRES) method, well adapted for large and sparse problems after preconditioning of the system \citep{Saad-Schultz-GMRES}. 
\subsection{Preliminary results}
\begin{figure}[ht!]
\centering
\includegraphics[width=\textwidth]{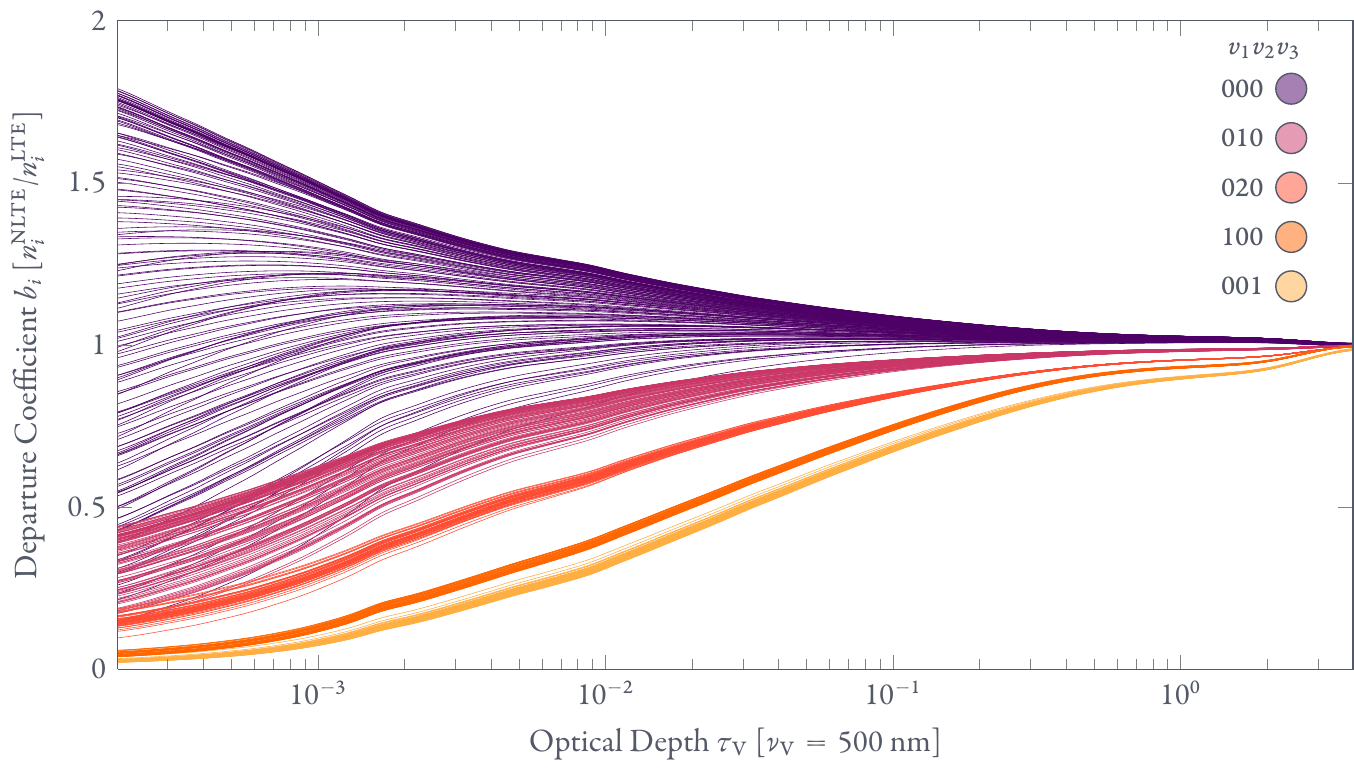}
\caption{Departure coefficients in a RSG model atmosphere. Vibrational bands are coded with different colours.}
\label{departure}
\end{figure}

The departure coefficients in a Betelgeuse-like model atmosphere are displayed in Fig. \ref{departure}. As expected from the 
critical densities, the strongest deviations from LTE are met in the upper atmosphere. Furthermore, the deviation from LTE is all 
larger when the energy is higher. The dispersion of the departure coefficients within a given vibrational level is remarkably low for highly-excited vibrational bands. In other words, these bands are under-populated with respect to LTE, but rotational levels are in relative LTE within these bands. This partially validates the potential use of the super-level technique for water  \citep[see][]{schweitzer2000}.

\begin{figure}[ht!]
\centering
\includegraphics[width=\textwidth]{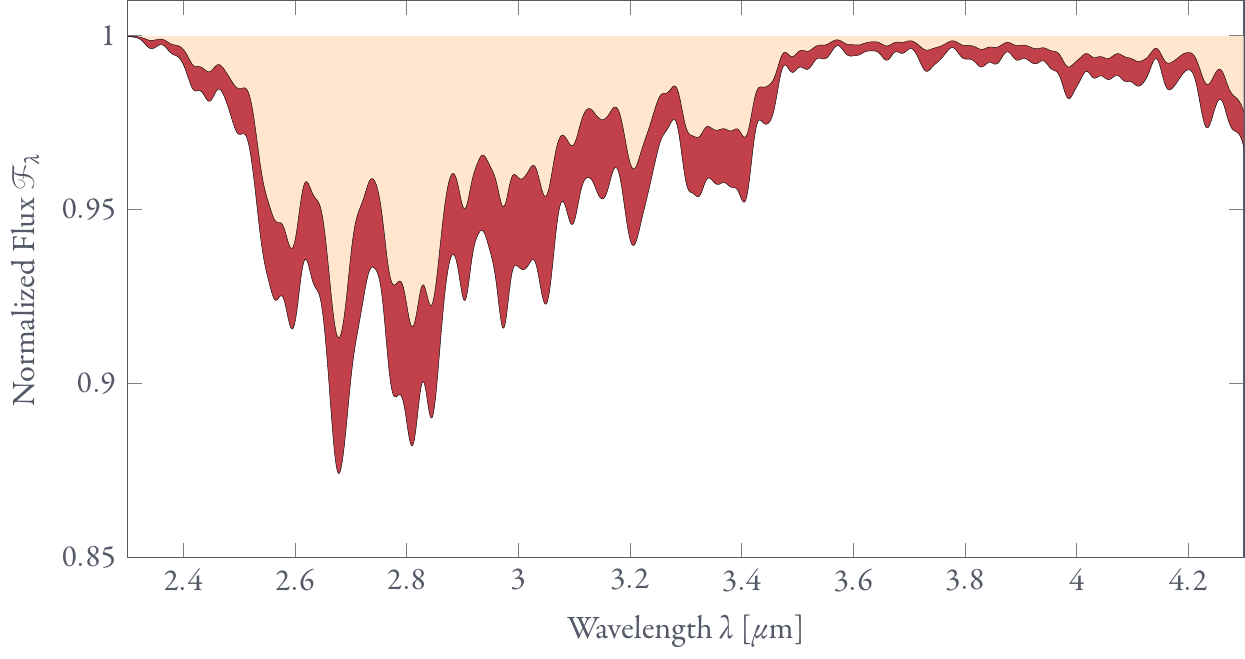}
\caption{Sample NLTE (red) and LTE (pink) water spectra emerging from a RSG.}
\label{spectrebr2}
\end{figure}

As shown in Fig. \ref{spectrebr2} the emerging spectrum exhibits water bands which are remarkably deeper compared to LTE predictions. NLTE seems thus to qualitatively mimic a MOLsphere. 

\section{Conclusions and perspectives}

Our preliminary results show an important deviation from LTE in water lines, which should be taken into account in the interpretation of infrared spectra of RSG and especially when discussing the MOLsphere hypothesis. These NLTE effects may also have strong implications on the thermodynamic structure of RSG atmospheres. For example, preliminary calculations show an increase by up to a factor of 4 in the cooling due to water molecules. The structure and the extent of these atmospheres may thus be strongly affected. For example, a shift outward of the contribution function peak is possible,  which would impact the interpretation of interferometric observations. Finally, radiative pressure on molecules may play a role in the mass loss process \citep[see][]{josselin2007}. 
%


\bibliography{lambert}
\bibliographystyle{astron}

\end{document}